\pdfoutput=1

\documentclass[12pt,a4paper]{article}

\usepackage{ifthen} 
\newboolean{pdflatex}
\setboolean{pdflatex}{true} 

\newboolean{articletitles}
\setboolean{articletitles}{true} 

\newboolean{uprightparticles}
\setboolean{uprightparticles}{false} 

\usepackage{microtype}
\usepackage{lineno}  % for line numbering during review
\usepackage{xspace} % To avoid problems with missing or double spaces after
\usepackage{graphicx}  % to include figures (can also use other packages)
\usepackage{color}
\usepackage{colortbl}
\graphicspath{{./figs/}} % Make Latex search fig subdir for figures

\usepackage{amsmath} % Adds a large collection of math symbols
\usepackage{amssymb}
\usepackage{amsfonts}
\usepackage{upgreek} % Adds in support for greek letters in roman typeset

\newcommand*\patchAmsMathEnvironmentForLineno[1]{%
\expandafter\let\csname old#1\expandafter\endcsname\csname #1\endcsname
\expandafter\let\csname oldend#1\expandafter\endcsname\csname
end#1\endcsname
 \renewenvironment{#1}%
   {\linenomath\csname old#1\endcsname}%
   {\csname oldend#1\endcsname\endlinenomath}%
}
\newcommand*\patchBothAmsMathEnvironmentsForLineno[1]{%
  \patchAmsMathEnvironmentForLineno{#1}%
  \patchAmsMathEnvironmentForLineno{#1*}%
}
\AtBeginDocument{%
\patchBothAmsMathEnvironmentsForLineno{equation}%
\patchBothAmsMathEnvironmentsForLineno{align}%
\patchBothAmsMathEnvironmentsForLineno{flalign}%
\patchBothAmsMathEnvironmentsForLineno{alignat}%
\patchBothAmsMathEnvironmentsForLineno{gather}%
\patchBothAmsMathEnvironmentsForLineno{multline}%
}

\usepackage{hyperref}    % Hyperlinks in references
\usepackage[all]{hypcap} % Internal hyperlinks to floats.

\def\lhcb {\mbox{LHCb}\xspace}
\def\ux85 {\mbox{UX85}\xspace}

\def\babar  {\mbox{BaBar}\xspace}
\def\belle  {\mbox{Belle}\xspace}

\ifthenelse{\boolean{uprightparticles}}%
{

 \def\Pmu         {\ensuremath{\upmu}\xspace}

 \def\Ppi         {\ensuremath{\uppi}\xspace}

 \def\Ppsi        {\ensuremath{\uppsi}\xspace}

 \def\PDelta      {\ensuremath{\Delta}\xspace}                 
 \def\PXi      {\ensuremath{\Xi}\xspace}                 
 \def\PLambda      {\ensuremath{\Lambda}\xspace}                 
 \def\PSigma      {\ensuremath{\Sigma}\xspace}                 
 \def\POmega      {\ensuremath{\Omega}\xspace}                 
 \def\PUpsilon      {\ensuremath{\Upsilon}\xspace}                 
 
 \def\PB      {\ensuremath{\mathrm{B}}\xspace}                 
                  
 \def\PD      {\ensuremath{\mathrm{D}}\xspace}

 \def\PJ      {\ensuremath{\mathrm{J}}\xspace}                 
 \def\PK      {\ensuremath{\mathrm{K}}\xspace}

 \def\Pc      {\ensuremath{\mathrm{c}}\xspace}

 \def\Pi      {\ensuremath{\mathrm{i}}\xspace}

 \def\Ps      {\ensuremath{\mathrm{s}}\xspace}

}
{

 \def\Pmu         {\ensuremath{\mu}\xspace}

 \def\Ppi         {\ensuremath{\pi}\xspace}

 \def\Ppsi        {\ensuremath{\psi}\xspace}                 
                  
 \mathchardef\PDelta="7101
 \mathchardef\PXi="7104
 \mathchardef\PLambda="7103
 \mathchardef\PSigma="7106
 \mathchardef\POmega="710A
 \mathchardef\PUpsilon="7107
                  
 \def\PB      {\ensuremath{B}\xspace}                 
                  
 \def\PD      {\ensuremath{D}\xspace}

 \def\PJ      {\ensuremath{J}\xspace}                 
 \def\PK      {\ensuremath{K}\xspace}

 \def\Pc      {\ensuremath{c}\xspace}

 \def\Pi      {\ensuremath{i}\xspace}

 \def\Ps      {\ensuremath{s}\xspace}

}

\def\mup        {\ensuremath{\Pmu^+}\xspace}
\def\mun        {\ensuremath{\Pmu^-}\xspace} % muon negative (\mum is taken)
\def\mumu       {\ensuremath{\Pmu^+\Pmu^-}\xspace}

\def\squark    {\ensuremath{\Ps}\xspace}

\def\cquark    {\ensuremath{\Pc}\xspace}
\def\cquarkbar {\ensuremath{\overline \cquark}\xspace}
\def\ccbar     {\ensuremath{\cquark\cquarkbar}\xspace}

\def\pion  {\ensuremath{\Ppi}\xspace}

\def\pip   {\ensuremath{\pion^+}\xspace}
\def\pim   {\ensuremath{\pion^-}\xspace}

\def\kaon  {\ensuremath{\PK}\xspace}
  \def\Kbar  {\kern 0.2em\overline{\kern -0.2em \PK}{}\xspace}

\def\Kz    {\ensuremath{\kaon^0}\xspace}
\def\Kzb   {\ensuremath{\Kbar^0}\xspace}
\def\KzKzb {\ensuremath{\Kz \kern -0.16em \Kzb}\xspace}
\def\Kp    {\ensuremath{\kaon^+}\xspace}
\def\Km    {\ensuremath{\kaon^-}\xspace}

\def\KpKm  {\ensuremath{\Kp \kern -0.16em \Km}\xspace}

\def\Kstarz  {\ensuremath{\kaon^{*0}}\xspace}

  \def\Dbar    {\kern 0.2em\overline{\kern -0.2em \PD}{}\xspace}
\def\D       {\ensuremath{\PD}\xspace}

\def\Dz      {\ensuremath{\D^0}\xspace}
\def\Dzb     {\ensuremath{\Dbar^0}\xspace}
\def\DzDzb   {\ensuremath{\Dz {\kern -0.16em \Dzb}}\xspace}
\def\Dp      {\ensuremath{\D^+}\xspace}
\def\Dm      {\ensuremath{\D^-}\xspace}

\def\DpDm    {\ensuremath{\Dp {\kern -0.16em \Dm}}\xspace}

\def\Dstarp  {\ensuremath{\D^{*+}}\xspace}

\def\B       {\ensuremath{\PB}\xspace}
  \def\Bbar    {\kern 0.18em\overline{\kern -0.18em \PB}{}\xspace}

\def\Bz      {\ensuremath{\B^0}\xspace}

\def\Bu      {\ensuremath{\B^+}\xspace}
\def\Bub     {\ensuremath{\B^-}\xspace}
\def\Bp      {\ensuremath{\Bu}\xspace}
\def\Bm      {\ensuremath{\Bub}\xspace}

\def\Bs      {\ensuremath{\B^0_\squark}\xspace}

\def\jpsi     {\ensuremath{{\PJ\mskip -3mu/\mskip -2mu\Ppsi\mskip 2mu}}\xspace}
\def\psitwos  {\ensuremath{\Ppsi{(2S)}}\xspace}

  \def\Y#1S{\ensuremath{\PUpsilon{(#1S)}}\xspace}% no space before {...}!

\def\Lbar {\ensuremath{\kern 0.1em\overline{\kern -0.1em\PLambda}}\xspace}

\def\BF         {{\ensuremath{\cal B}\xspace}}

\newcommand{\decay}[2]{\ensuremath{#1\!\to #2}\xspace}         % {\Pa}{\Pb \Pc}

\def\to                 {\ensuremath{\rightarrow}\xspace}

\def\qsq       {\ensuremath{q^2}\xspace}

\def\AT#1     {\ensuremath{A_{\mathrm{T}}^{#1}}\xspace}           % 2

\def\ctl       {\ensuremath{\cos{\theta_l}}\xspace}

\def\C#1      {\ensuremath{\mathcal{C}_{#1}}\xspace}                       % 9
\def\Cp#1     {\ensuremath{\mathcal{C}_{#1}^{'}}\xspace}                    % 7
\def\Ceff#1   {\ensuremath{\mathcal{C}_{#1}^{\mathrm{(eff)}}}\xspace}        % 9  
\def\Cpeff#1  {\ensuremath{\mathcal{C}_{#1}^{'\mathrm{(eff)}}}\xspace}       % 7
\def\Ope#1    {\ensuremath{\mathcal{O}_{#1}}\xspace}                       % 2
\def\Opep#1   {\ensuremath{\mathcal{O}_{#1}^{'}}\xspace}                    % 7

\newcommand{\tev}{\ensuremath{\mathrm{\,Te\kern -0.1em V}}\xspace}
\newcommand{\gev}{\ensuremath{\mathrm{\,Ge\kern -0.1em V}}\xspace}
\newcommand{\mev}{\ensuremath{\mathrm{\,Me\kern -0.1em V}}\xspace}
\newcommand{\kev}{\ensuremath{\mathrm{\,ke\kern -0.1em V}}\xspace}
\newcommand{\ev}{\ensuremath{\mathrm{\,e\kern -0.1em V}}\xspace}
\newcommand{\gevc}{\ensuremath{{\mathrm{\,Ge\kern -0.1em V\!/}c}}\xspace}
\newcommand{\mevc}{\ensuremath{{\mathrm{\,Me\kern -0.1em V\!/}c}}\xspace}
\newcommand{\gevcc}{\ensuremath{{\mathrm{\,Ge\kern -0.1em V\!/}c^2}}\xspace}
\newcommand{\gevgevcccc}{\ensuremath{{\mathrm{\,Ge\kern -0.1em V^2\!/}c^4}}\xspace}
\newcommand{\mevcc}{\ensuremath{{\mathrm{\,Me\kern -0.1em V\!/}c^2}}\xspace}

\def\mum  {\ensuremath{\,\upmu\rm m}\xspace}

\def\invfb   {\ensuremath{\mbox{\,fb}^{-1}}\xspace}

\def\deriv {\ensuremath{\mathrm{d}}}

\def\gsim{{~\raise.15em\hbox{$>$}\kern-.85em
          \lower.35em\hbox{$\sim$}~}\xspace}
\def\lsim{{~\raise.15em\hbox{$<$}\kern-.85em
          \lower.35em\hbox{$\sim$}~}\xspace}

\def\pt         {\mbox{$p_{\rm T}$}\xspace}

\def\evtgen     {\mbox{\textsc{EvtGen}}\xspace}
\def\pythia     {\mbox{\textsc{Pythia}}\xspace}

\def\geant      {\mbox{\textsc{Geant4}}\xspace}

\def\photos     {\mbox{\textsc{Photos}}\xspace}

\def\tell1  {TELL1\xspace}
\def\ukl1   {UKL1\xspace}

\usepackage{cite} % Allows for ranges in citations
\usepackage{mciteplus}

\begin{document}

\renewcommand{\thefootnote}{\fnsymbol{footnote}}
\setcounter{footnote}{1}

\begin{titlepage}
\pagenumbering{roman}

\vspace*{-1.5cm}
\centerline{\large EUROPEAN ORGANIZATION FOR NUCLEAR RESEARCH (CERN)}
\vspace*{1.5cm}
\hspace*{-0.5cm}
\begin{tabular*}{\linewidth}{lc@{\extracolsep{\fill}}r}
\ifthenelse{\boolean{pdflatex}}
{\vspace*{-2.7cm}\mbox{\!\!\!\includegraphics[width=.14\textwidth]{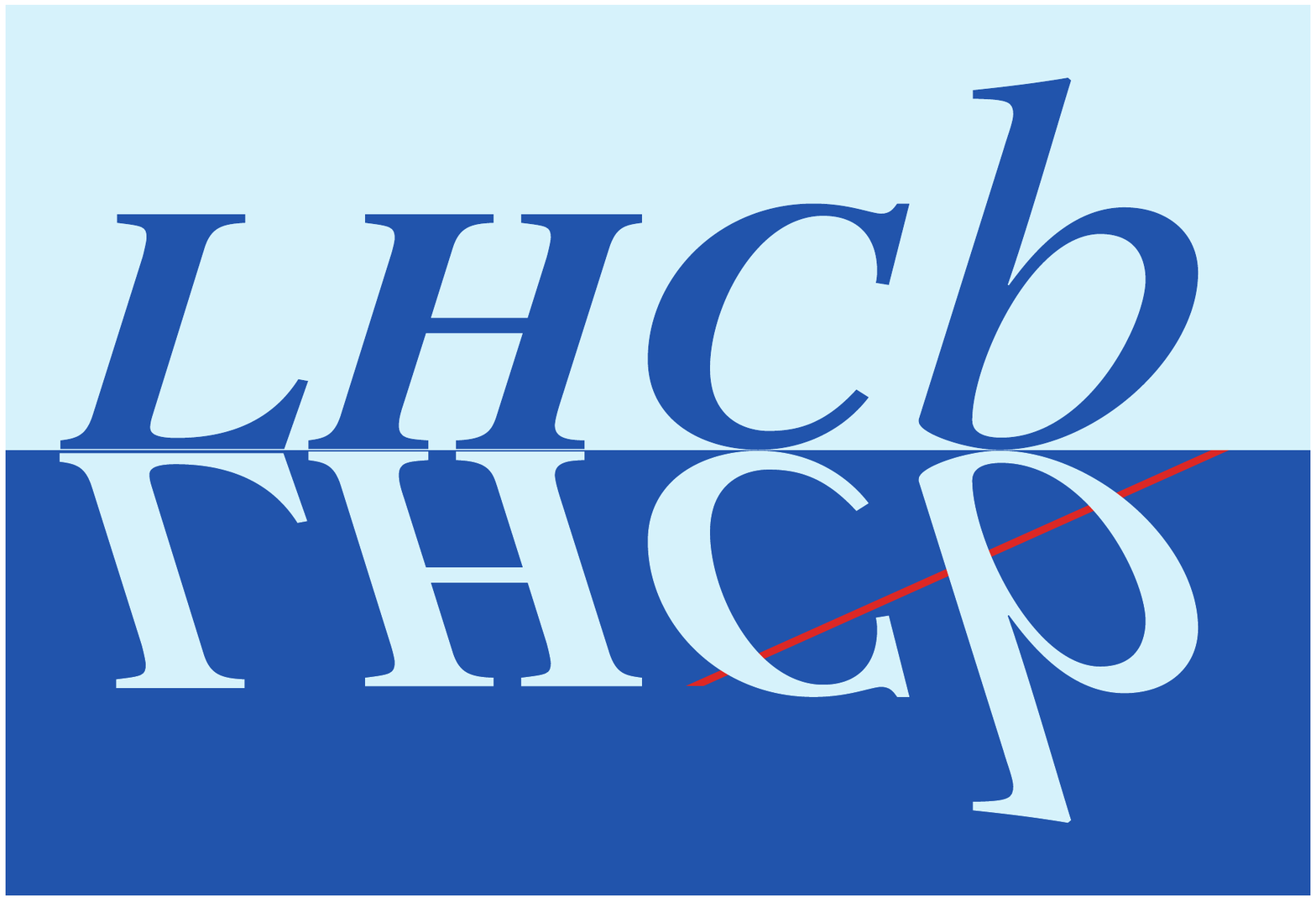}} & &}%
{\vspace*{-1.2cm}\mbox{\!\!\!\includegraphics[width=.12\textwidth]{figs/lhcb-logo.eps}} & &}%
\\
 & & CERN-PH-EP-2012-263 \\  % ID 
 & & LHCb-PAPER-2012-024 \\  % ID 
 & & 19 September 2012   \\  % \today 
 & & \\
\end{tabular*}

\vspace*{4.0cm}

{\bf\boldmath\huge
\begin{center}
  Differential branching fraction \\ and angular analysis of \\ the \decay{\Bp}{\Kp\mumu} decay
\end{center}
}

\vspace*{2.0cm}

\begin{center}
The LHCb collaboration\footnote{Authors are listed on the following pages.}
\end{center}

\vspace{\fill}

\begin{abstract}
  \noindent 
The angular distribution and differential branching fraction of the decay ${\decay{\Bp}{\Kp\mumu}}$ are studied with a dataset corresponding to 1.0\invfb of integrated luminosity, collected by the LHCb experiment. The angular distribution is measured in bins of dimuon invariant mass squared and found to be consistent with Standard Model expectations.  Integrating the differential branching fraction over the full dimuon invariant mass range yields a total branching fraction of $\BF(\decay{\Bp}{\Kp\mumu}) = (4.36 \pm 0.15 \pm 0.18)\times 10^{-7}$. These measurements are the most precise to date of the \decay{\Bp}{\Kp\mumu} decay.
\end{abstract}

\vspace*{2.0cm}

\begin{center}
  To be submitted to Journal of High Energy Physics
\end{center}

\vspace{\fill}

\end{titlepage}

\newpage
\setcounter{page}{2}
\mbox{~}
\newpage

%%%%%%%%%%%%%%%%%%%%%%%%%%%%%%%%%%%%%%%%%%
\centerline{\large\bf LHCb collaboration}
\begin{flushleft}
\small
R.~Aaij$^{38}$, 
C.~Abellan~Beteta$^{33,n}$, 
A.~Adametz$^{11}$, 
B.~Adeva$^{34}$, 
M.~Adinolfi$^{43}$, 
C.~Adrover$^{6}$, 
A.~Affolder$^{49}$, 
Z.~Ajaltouni$^{5}$, 
J.~Albrecht$^{35}$, 
F.~Alessio$^{35}$
% ...

{\footnotesize \it
% ...
$ ^{5}$Clermont Universit\'{e}, Universit\'{e} Blaise Pascal, CNRS/IN2P3, LPC, Clermont-Ferrand, France\\
$ ^{6}$CPPM, Aix-Marseille Universit\'{e}, CNRS/IN2P3, Marseille, France\\
% ...
$ ^{11}$Physikalisches Institut, Ruprecht-Karls-Universit\"{a}t Heidelberg, Heidelberg, Germany\\
% ...
$ ^{33}$Universitat de Barcelona, Barcelona, Spain\\
$ ^{34}$Universidad de Santiago de Compostela, Santiago de Compostela, Spain\\
$ ^{35}$European Organization for Nuclear Research (CERN), Geneva, Switzerland\\
% ...
$ ^{38}$Nikhef National Institute for Subatomic Physics, Amsterdam, The Netherlands\\
% ...
$ ^{43}$H.H. Wills Physics Laboratory, University of Bristol, Bristol, United Kingdom\\
% ...
$ ^{49}$Oliver Lodge Laboratory, University of Liverpool, Liverpool, United Kingdom\\
% ...
$ ^{n}$LIFAELS, La Salle, Universitat Ramon Llull, Barcelona, Spain\\
% ...
}
\end{flushleft}
%%%%%%%%%%%%%%%%%%%%%%%%%%%%%%%%%%%%%%%%%%

\cleardoublepage

\renewcommand{\thefootnote}{\arabic{footnote}}
\setcounter{footnote}{0}

\pagestyle{plain} 
\setcounter{page}{1}
\pagenumbering{arabic}

%\linenumbers

\section{Introduction}
\label{sec:introduction}

The \decay{\Bp}{\Kp\mumu} decay\footnote{Charge conjugation is implied throughout this paper unless explicitly stated otherwise.} is a $b \to s$ flavour changing neutral current process that is mediated in the Standard Model (SM) by electroweak box and penguin diagrams. In many well motivated extensions to the SM, new particles can enter in competing loop diagrams, modifying the branching fraction of the decay or the angular distribution of the dimuon system. The differential decay rate of the \Bp (\Bm) decay, as a function of $\cos\theta_\ell$, the cosine of the angle between the \mun (\mup) and the \Kp (\Km) in the rest frame of the dimuon system, can be written as 

\begin{equation}
  \label{eq:angular}
  \frac{1}{\Gamma} \frac{\deriv\Gamma [\Bp \to \Kp \mumu]}{\deriv\!\ctl}
    = \frac{3}{4} (1 - F_{\rm H}) (1 - \cos^2\theta_l) 
    + \frac{1}{2} F_{\rm H} + A_{\rm FB} \ctl ~,
\end{equation} 

\noindent which depends on two parameters, the forward-backward asymmetry of the dimuon system, $A_{\rm FB}$, and the parameter $F_{\rm H}$~\cite{Ali:1999mm,Bobeth:2007dw}. If muons were massless, $F_{\rm H}$ would be proportional to the contributions from (pseudo-)scalar and tensor operators to the partial width, $\Gamma$. The partial width, $A_{\rm FB}$ and $F_{\rm H}$ are functions of the dimuon invariant mass squared ($\qsq = m_{\mumu}^{2}$). 

In contrast to the case of the \decay{\Bz}{\Kstarz\mumu}~\cite{Aaij:2011aa,*LHCb-CONF-2012-008} decay, $A_{\rm FB}$ is zero for \decay{\Bp}{\Kp\mumu} in the SM. Any non-zero value for $A_{\rm FB}$ would point to a contribution from new particles that would extend the set of SM operators. In models with (pseudo-)scalar or tensor-like couplings $|A_{\rm FB}|$ can be enhanced by up to 15\%~\cite{Alok:2008wp, Bobeth:2007dw}. Similarly, $F_{\rm H}$ is close to zero in the SM (see Fig.~\ref{fig:angular}), but can be enhanced in new physics models, with (pseudo-)scalar or tensor-like couplings, up to $F_{\rm H} \lsim 0.5$. Recent predictions for these parameters in the SM are described in Refs.~\cite{Bobeth:2007dw, Khodjamirian:2010vf, Bobeth:2011nj}. Any physics model has to satisfy the constraint $|A_{\rm FB}| \leq F_{\rm H} / 2$ for Eq.~(\ref{eq:angular}) to stay positive in all regions of phase space. The contributions of scalar and pseudoscalar operators to $A_{\rm FB}$ and $F_{\rm H}$ are constrained by recent limits on the branching fraction of \decay{\Bs}{\mumu}~\cite{Aaij:2012ac, *Chatrchyan:2012rg}. The differential branching fraction of \decay{\Bp}{\Kp\mumu} can be used to constrain the contributions from (axial-)vector couplings in the SM operator basis~\cite{Altmannshofer:2011gn, Bobeth:2011nj, Altmannshofer:2012az}. 

The relative decay rate of \decay{\Bp}{\Kp\mumu} to \decay{\Bz}{\Kz\mumu} has previously been studied by the \lhcb collaboration in the context of a measurement of the isospin asymmetry~\cite{Aaij:1446187}. This paper presents a measurement of the differential branching fraction ($\deriv\BF/\deriv\qsq$), $F_{\rm H}$ and $A_{\rm FB}$ of the decay \decay{\Bp}{\Kp\mumu} in seven bins of \qsq and a measurement of the total branching fraction. The analysis is based on 1.0\invfb of integrated luminosity collected in $\sqrt{s} = 7\tev$ $pp$ collisions by the \lhcb experiment in 2011.

\section{Experimental setup}
\label{sec:detector}

The \lhcb detector~\cite{Alves:2008zz} is a single-arm forward spectrometer, covering the \mbox{pseudorapidity} range $2 < \eta < 5$, that is designed to study $b$ and $c$ hadron decays.  A dipole magnet with a bending power of 4\,Tm and a large area tracking detector provide a momentum resolution ranging from 0.4\% for tracks with a momentum of 5\gevc to 0.6\% for a momentum of 100\gevc. A silicon micro-strip detector, located around the $pp$ interation region, provides excellent separation of \B meson decay vertices from the primary $pp$ interaction and an impact parameter resolution of 20\mum for tracks with high transverse \mbox{momentum (\pt)}. Two ring-imaging Cherenkov (RICH) detectors provide kaon-pion separation in the momentum range $2-100\gevc$. Muons are identified based on hits created in a system of multiwire proportional chambers interleaved with iron filters. The LHCb trigger comprises a hardware trigger and a two-stage software trigger that performs a full event reconstruction.

Samples of simulated events are used to estimate the contribution from specific sources of exclusive backgrounds and the efficiency to trigger, reconstruct and select the \decay{\Bp}{\Kp\mumu} signal. The simulated $pp$ interactions are generated using
\pythia~6.4~\cite{Sjostrand:2006za} with a specific \lhcb
configuration~\cite{LHCb-PROC-2010-056}.  Decays of hadronic particles
are then described by \evtgen~\cite{Lange:2001uf} in which final state
radiation is generated using \photos~\cite{Golonka:2005pn}. Finally, the \geant toolkit~\cite{Allison:2006ve, *Agostinelli:2002hh} is used to simulate the detector response to the particles produced by  \pythia/\evtgen, as described in
Ref.~\cite{LHCb-PROC-2011-006}.  The simulated samples are corrected for differences between data and simulation in the \Bp momentum spectrum, the detector impact parameter resolution, particle identification and tracking system performance.

\section{Selection of signal candidates}

The \decay{\Bp}{\Kp\mumu} candidates are selected from events that have been triggered by a single high transverse-momentum muon, with $\pt > 1.5\gevc$, in the hardware trigger. In the first stage of the software trigger, candidates are selected if there is a reconstructed track in the event with high impact parameter with respect to the primary $pp$ interaction and high \pt~\cite{LHCb-PUB-2011-003}. In the second stage of the software trigger, candidates are triggered on the kinematic properties of the partially or fully reconstructed \Bp candidate~\cite{LHCb-PUB-2011-016}. 

Signal candidates are then selected for further analysis based on the following requirements: the \Bp decay vertex is separated from the primary $pp$ interaction; the \Bp candidate impact parameter is small, and the kaon and muon impact parameters large, with respect to the primary $pp$ interaction; the \Bp candidate momentum vector points along the \Bp line of flight to one of the primary $pp$ interactions in the event.

A tighter multivariate selection, using a Boosted Decision Tree (BDT)~\cite{Breiman} with the AdaBoost algorithm\cite{AdaBoost}, is then applied to select a clean sample of \decay{\Bp}{\Kp\mumu} candidates.  The BDT uses kinematic variables including the reconstructed \Bp decay time, the angle between the \Bp line of flight and the \Bp momentum vector, the quality of the vertex fit of the reconstructed \Bp candidate, impact parameter (with respect to the primary $pp$ interaction) and \pt of the \Bp and muons and the track quality of the kaon. The  variables that are used in the BDT provide good separating power between signal and background, while minimising acceptance effects in $q^{2}$ and $\cos\theta_{\ell}$ that could bias the differential branching fraction, $A_{\rm FB}(\qsq)$ or $F_{\rm H}(\qsq)$.   The multivariate selection is trained on data, using \decay{\Bp}{\Kp\jpsi} (\decay{\jpsi}{\mumu}) candidates as a proxy for the signal and \decay{\Bp}{\Kp\mumu} candidates from the upper mass sideband ($5350 < m_{\Kp\mumu} < 5600\mevcc$) for the background. The training and testing of the BDT is carried out using a data sample corresponding to 0.1\invfb of integrated luminosity, that is not used in the subsequent analysis. The BDT selection is $85-90\%$ (depending on \qsq) efficient on simulated candidates that have passed the earlier selection. 

Finally, a neural network, using information from the RICH~\cite{Forty:1999sg}, calorimeters and muon system is used to reject backgrounds where a pion is incorrectly identified as the kaon from the \decay{\Bp}{\Kp\mumu} decay. The network is trained on simulated event samples to give the posterior probability for charged hadrons to be correctly identified. The particle identification performance of the network is calibrated using pions and kaons from the decay chain \decay{\Dstarp}{\Dz (\to \Km\pip) \pip} in the data. Based on simulation, the efficiency of the neural network particle identification requirement is estimated to be $\gsim 95\%$ on the signal.

The contribution from combinatorial backgrounds, where the reconstructed \Kp, \mup and \mun do not come from the same $b$-hadron decay, is reduced to a small level by the multivariate selection. Remaining backgrounds come from exclusive $b$-hadron decays. The decays \decay{\Bp}{\Kp\jpsi} and \decay{\Bp}{\Kp\psitwos} are rejected by removing the regions of dimuon invariant  mass around the charmonium resonances ($2946 < m_{\mumu} < 3176\mevcc$ and $3586 < m_{\mumu} < 3776\mevcc$). Candidates with $m_{\Kp\mumu} < 5170\mevcc$ were also removed to reject backgrounds from partially reconstructed $B$ decays, such as \decay{\Bz}{\Kstarz\mumu}. The potential background from \decay{\Bp}{\Kp\jpsi} (\decay{\jpsi}{\mup\mun}), where the kaon is identified as a muon and a muon as the kaon, is reduced by requiring that the kaon candidate fails the muon identification criteria if the $\Kp\mun$ mass is consistent with that of the \jpsi or \psitwos. Candidates with a $\Kp\mun$ mass consistent with coming from a misidentified \decay{\Dzb}{\Kp\pim} decay are rejected to remove contributions from \decay{\Bp}{\Dzb\pip}. After the application of all of the selection criteria, the dominant sources of exclusive background are \decay{\Bp}{\Kp\pim\pip}~\cite{Aubert:2008bj} and \decay{\Bp}{\pip\mumu}~\cite{LHCb-PAPER-2012-020,*LHCb-CONF-2012-006}. These are determined from simulation to be at the level of $(1.5 \pm 0.7)\%$ and $(1.2 \pm 0.2)\%$  of the signal, respectively.

\section{Differential and total branching fraction} 
\label{sec:diffbr}

\begin{figure}
\centering
\includegraphics[scale=0.52]{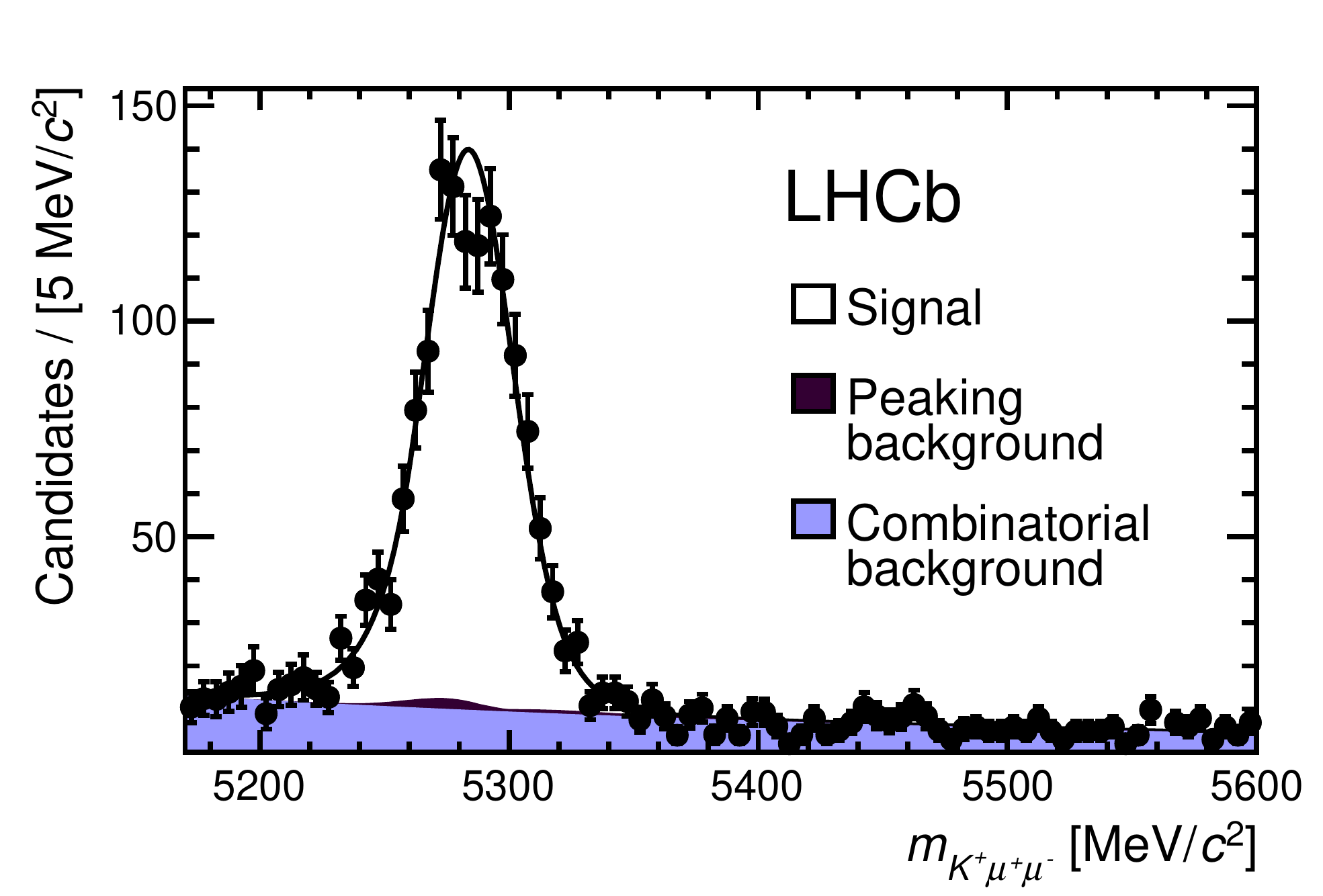}
\caption{Invariant mass of selected \decay{\Bp}{\Kp\mumu} candidates with $0.05 < q^{2} < 22.00\gev^{2}/c^{4}$. Candidates with a dimuon invariant mass consistent with that of the \jpsi or \psitwos are excluded. The peaking background contribution from the decays \decay{\Bp}{\Kp\pip\pim} and \decay{\Bp}{\pip\mumu} is indicated in the figure. \label{fig:mass}}
\end{figure}

The $\Kp\mumu$ invariant mass distribution of the selected \decay{\Bp}{\Kp\mumu} candidates is shown in Fig.~\ref{fig:mass}. The number of signal candidates is estimated by performing an extended unbinned maximum likelihood fit to the $\Kp\mumu$ invariant mass distribution of the selected candidates. The signal line-shape is extracted from a fit to a \decay{\Bp}{\Kp\jpsi} (\decay{\jpsi}{\mumu}) control sample (which is two orders of magnitude larger than the signal sample), and is parameterised by the sum of two Crystal Ball functions~\cite{Skwarnicki:1986xj}. The combinatorial background is parameterised by a slowly falling exponential distribution. Contributions from \decay{\Bp}{\Kp\pip\pim} and \decay{\Bp}{\pip\mumu} decays are included in the fit. The line shapes of these peaking backgrounds are taken from simulated events. In total, $1232 \pm 40$ \decay{\Bp}{\Kp\mumu} signal candidates are observed in the $0.05 < q^{2} < 22.00\gev^{2}/c^{4}$ range. The yields in each of the \qsq bins used in the subsequent analysis are shown in Table~\ref{tab:results}.

\begin{figure}
\centering
\includegraphics[scale=0.52]{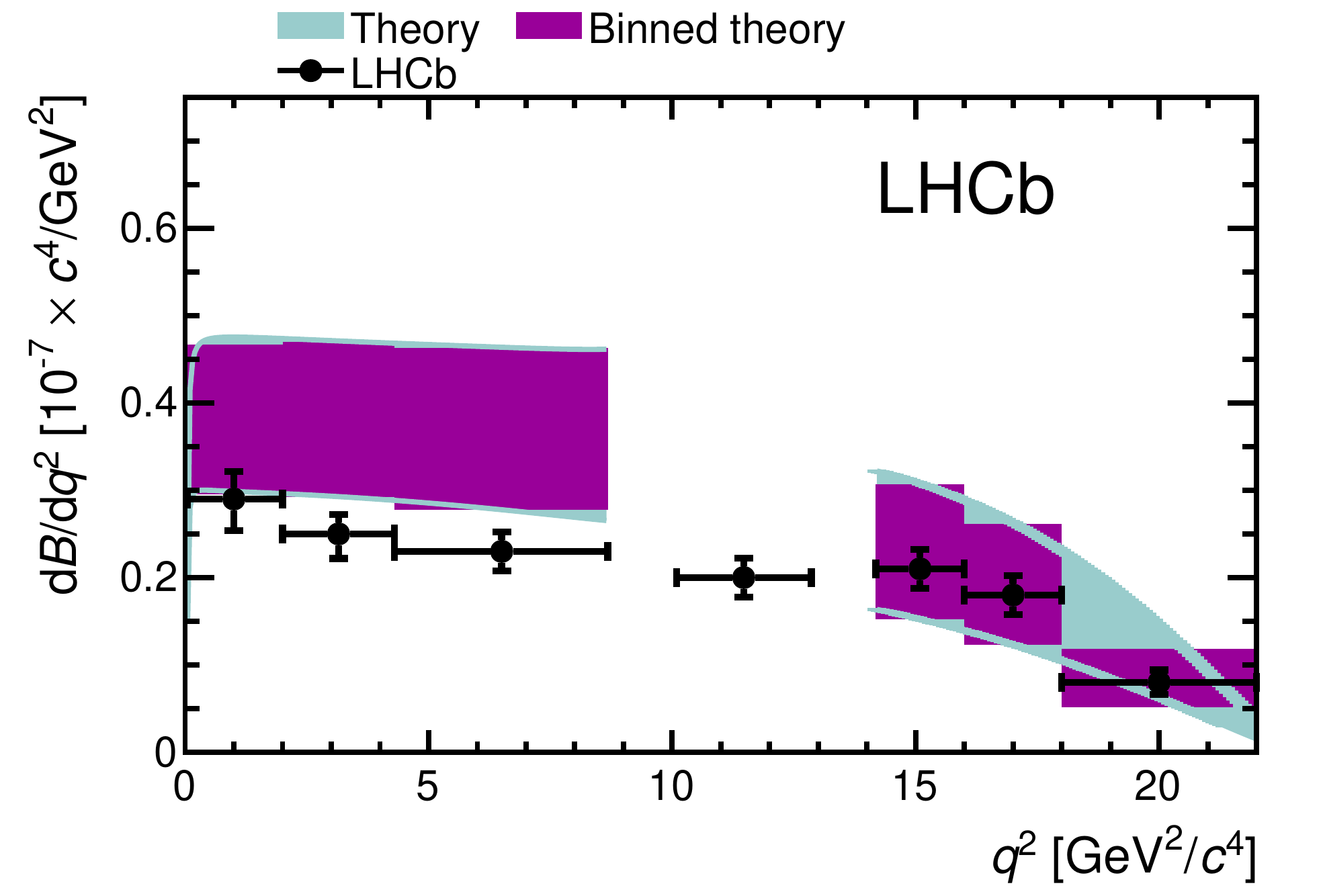}
\caption{Differential branching fraction of \decay{\Bp}{\Kp\mumu} as a function of the dimuon invariant mass squared, $q^{2}$. The SM theory prediction (see text) is given as the continuous cyan (light) band and the rate-average of this prediction across the $q^{2}$ bin is indicated by the purple (dark) region. No SM prediction is included for the regions close to the narrow \ccbar resonances. \label{fig:diffbr}}
\end{figure}

The differential branching fraction in each of the \qsq bins is estimated by normalising the \decay{\Bp}{\Kp\mumu} event yield, $N_{\text{sig}}$, in the \qsq bin to the total event yield of the \decay{\Bp}{\Kp\jpsi} sample, $N_{\Kp\jpsi}$, and correcting for the relative efficiency between the two decays in the \qsq bin, $\varepsilon_{\Kp\jpsi}/\varepsilon_{\Kp\mumu}$, 

\begin{equation}
\frac{\deriv \BF}{\deriv \qsq} = \frac{1}{\qsq_{\text{max}}-\qsq_{\text{min}}}\frac{N_{\text{sig}}}{N_{\Kp\jpsi}}\frac{\varepsilon_{\Kp\jpsi}}{\varepsilon_{\Kp\mumu}}\times\BF(\decay{\Bp}{\Kp\jpsi})\times\BF(\decay{\jpsi}{\mumu}) ~.
\end{equation}

\noindent The branching fractions of \decay{\Bp}{\Kp\jpsi} and \decay{\jpsi}{\mumu} are $\BF(\decay{\Bp}{\Kp\jpsi}) = (1.014 \pm 0.034) \times 10^{-3}$ and $\BF(\decay{\jpsi}{\mumu}) = (5.93 \pm 0.06) \times 10^{-2}$~\cite{PDG2012}. The resulting differential branching fraction is shown in Fig.~\ref{fig:diffbr}.

The bands shown in Fig.~\ref{fig:diffbr} indicate the theoretical prediction for the differential branching fraction and are calculated using input from Refs.~\cite{Bobeth:2011nj} and \cite{Bobeth:2011gi}. In the low $q^{2}$ region, the calculations are based on QCD factorisation and soft collinear effective theory (SCET)~\cite{Beneke:2001at}, which profit from  having a heavy \Bp meson and an energetic kaon. In the soft-recoil, high $q^{2}$ region, an operator product expansion (OPE) is used to estimate the long-distance contributions from quark loops~\cite{Grinstein:2004vb, Beylich:2011aq}.  No theory prediction is included in the region close to the narrow \ccbar resonances (the \jpsi and \psitwos) where the assumptions from QCD factorisation/SCET and the OPE break down. The form-factor calculations are taken from Ref.~\cite{Ball:2004ye}. A dimensional estimate is made on the uncertainty on the decay amplitudes from QCD factorisation/SCET~\cite{Egede:2008uy}.

The total branching fraction is measured to be 

\begin{displaymath}
{\BF(\decay{\Bp}{\Kp\mumu}) = ( 4.36 \pm 0.15 \pm 0.18 ) \times 10^{-7} ~,}
\end{displaymath}

\noindent by summing over the partial branching fractions and accounting for the \qsq regions that are not used in the differential branching fraction analysis. These regions account for $\sim14.3\%$ of the total branching fraction (no uncertainty is assigned to this number). This estimate ignores long distance effects and uses a model for $\deriv\Gamma/\deriv\qsq$ described in Ref.~\cite{Ali:1999mm} to extrapolate across the \ccbar resonance region. The values of the Wilson coefficients and the form-factors used in this model have been updated according to Refs.~\cite{Ali:2002jg} and \cite{Ball:2004ye}.

\section{Angular analysis}
\label{sec:angularanalysis}

In each bin of \qsq, $A_{\rm FB}$ and $F_{\rm H}$ are estimated by performing a simultaneous unbinned maximum likelihood fit to the $\Kp\mumu$ invariant mass and $\cos\theta_{\ell}$ distribution of the \Bp candidates.  The candidates are weighted to account for the effects of the detector reconstruction, trigger and the event selection. The weights are derived from a SM simulation of the \decay{\Bp}{\Kp\mumu} decay in bins of width $0.5\gev^{2}/c^{4}$ in \qsq and 0.1 in $\cos\theta_{\ell}$. This binning is investigated as a potential source of systematic uncertainty.  The largest weights (and largest acceptance effects) apply to events with extreme values of $\cos\theta_{\ell}$ ($|\cos\theta_{\ell}|\sim 1$) at low \qsq. This distortion arises mainly from the requirement for a muon to have $p \gsim 3\gevc$ to reach the LHCb muon system. This effect is well modelled in the simulation. 
 
Equation~(\ref{eq:angular}) is used to describe the signal angular distribution.  The background angular and mass shapes are treated as independent in the fit. The angular distribution of the background is parameterised by a second-order Chebychev polynomial, which is observed to describe well the background away from the signal mass window ($5230 < m_{\Kp\mumu} < 5330\mevcc$). 

\begin{figure}
\centering
\includegraphics[scale=0.395]{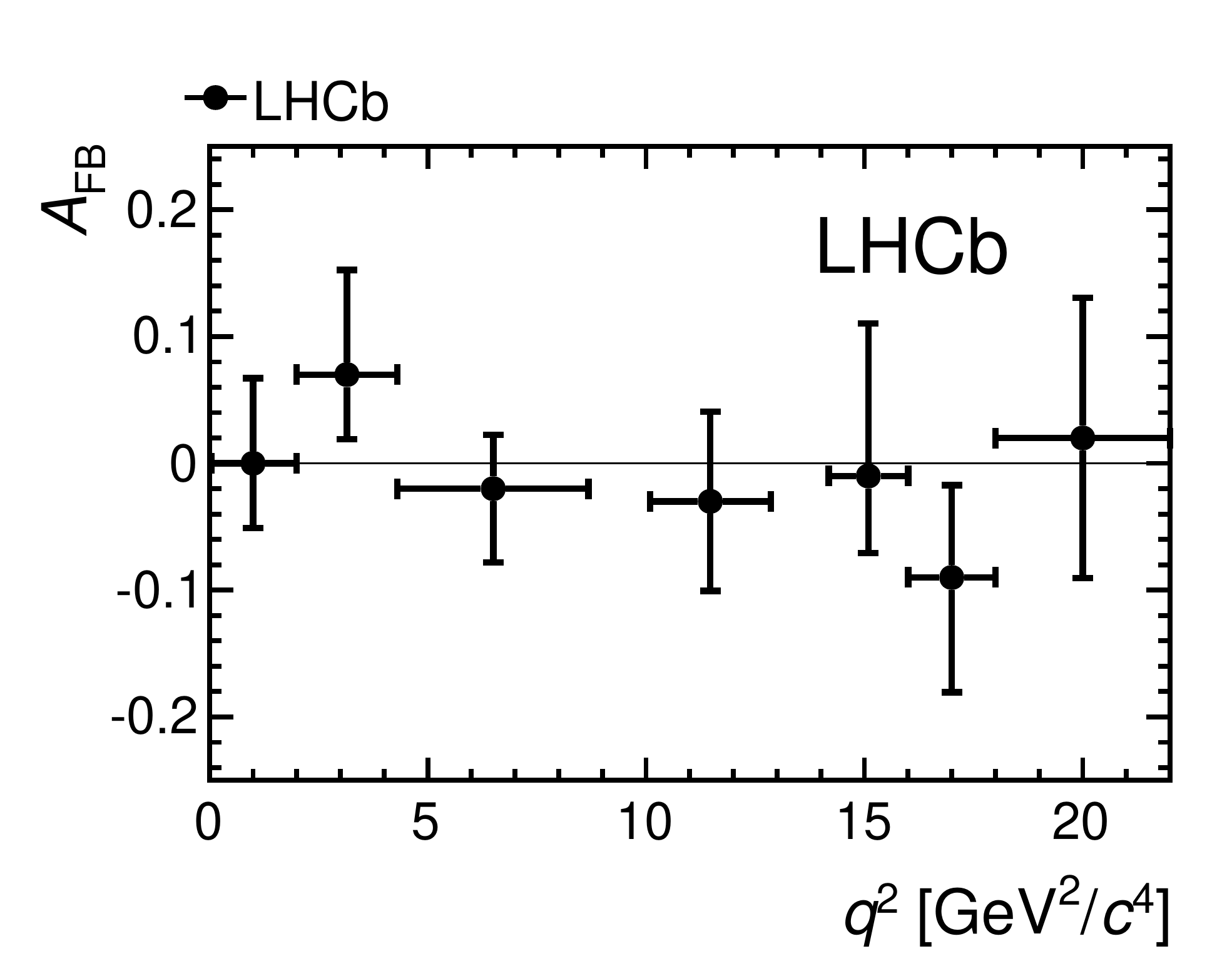}
\includegraphics[scale=0.395]{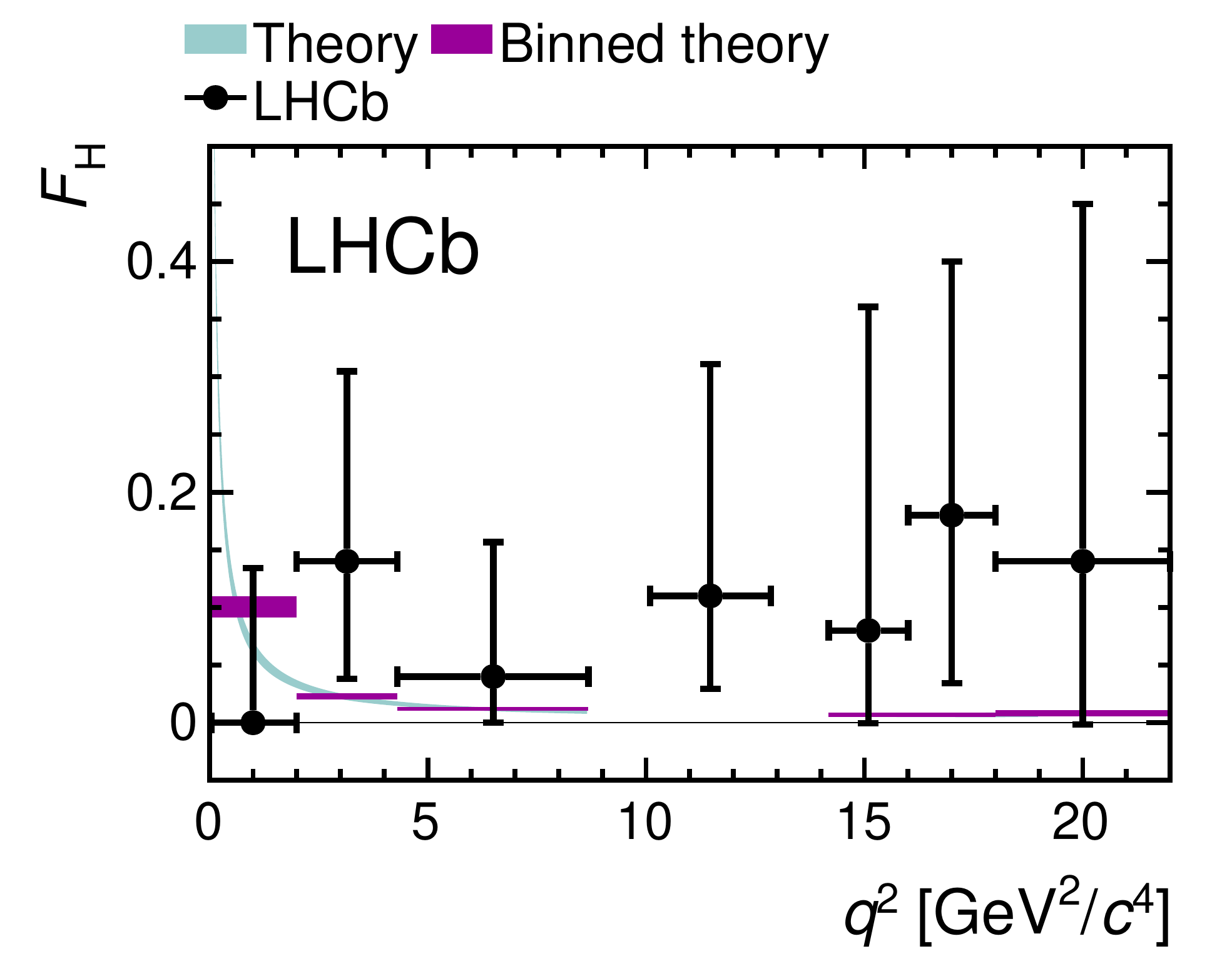}
\caption{Dimuon forward-backward asymmetry, $A_{\rm FB}$, and the parameter $F_{\rm H}$ for \decay{\Bp}{\Kp\mumu} as a function of the dimuon invariant mass squared, \qsq. The SM theory prediction (see text) for $F_{\rm H}$ is given as the continuous cyan (light) band and the rate-average of this prediction across the \qsq bin is indicated by the purple (dark) region. No SM prediction is included for the regions close to the narrow \ccbar resonances. \label{fig:angular}}
\end{figure}

The resulting values of $A_{\rm FB}$ and $F_{\rm H}$ in the bins of \qsq are indicated in Fig.~\ref{fig:angular} and in Table~\ref{tab:results}. The measured values of $A_{\rm FB}$ are consistent with the SM expectation of zero asymmetry. The 68\% confidence intervals on $A_{\rm FB}$ and $F_{\rm H}$ are estimated using pseudo-experiments and the Feldman-Cousins technique~\cite{Feldman:1997qc}. This avoids potential biases in the estimate of the parameter uncertainties that come from using event weights in the likelihood fit or from the boundary condition ($|A_{\rm FB}| \leq F_{\rm H} / 2$). When estimating the uncertainty on $A_{\rm FB}$ ($F_{\rm H}$), $F_{\rm H}$ ($A_{\rm FB}$) is treated as a nuisance parameter (along with the background parameters in the fit). The maximum-likelihood estimate of the nuisance parameters is used when generating the pseudo-experiments. The resulting confidence intervals ignore correlations between $A_{\rm FB}$ and $F_{\rm H}$ and are not simultaneously valid at the 68\% confidence level. 

\begin{table*}
\centering
\caption{Signal yield ($N_{\text{sig}}$), differential branching fraction ($\deriv\BF/\deriv\qsq$), the parameter $F_{\rm H}$ and dimuon forward-backward asymmetry ($A_{\rm FB}$) for the \decay{\Bp}{\Kp\mumu} decay in the \qsq bins used in the analysis. Results are also given in the $1 < \qsq < 6\gev^{2}/c^{4}$ range where theoretical uncertainties are best under control. \label{tab:results}}
\setlength{\extrarowheight}{2pt}
\resizebox{\linewidth}{!}{
\begin{tabular}{c|cccc}
$q^{2}$ $(\gev^{2}/c^{4})$ & $N_{\text{sig}}$ & $\deriv\BF/\deriv\qsq$ $(10^{-8} \gev^{-2} c^{4})$ & $F_{\rm H}$ & $A_{\rm FB}$ \\ 
\hline
$0.05 - 2.00$ & $159 \pm 14$ & $2.85 \pm 0.27 \pm 0.14$ & $0.00\,^{\,+0.12}_{\,-0.00}\,^{\,+0.06}_{\,-0.00}$ &  $\phantom{-}0.00\,^{\,+0.06}_{\,-0.05}\,^{\,+0.03}_{\,-0.01}$  \\
$2.00 - 4.30$ & $164\pm14$ & $2.49 \pm 0.23 \pm 0.10$ & $0.14\,^{\,+0.16}_{\,-0.10}\,^{\,+0.04}_{\,-0.02}$ &  $\phantom{-}0.07\,^{\,+0.08}_{\,-0.05}\,^{\,+0.02}_{\,-0.01}$\\
$4.30 - 8.68$ & $327\pm20$ & $2.29 \pm 0.16 \pm 0.09$ & $0.04\,^{\,+0.10}_{\,-0.04}\,^{\,+0.06}_{\,-0.04}$ &  $-0.02\,^{\,+0.03}_{\,-0.05}\,^{\,+0.03}_{\,-0.03}$\\
$10.09 - 12.86$  & $211\pm17$ & $2.04 \pm 0.18 \pm 0.08$ & $0.11\,^{\,+0.20}_{\,-0.08}\,^{\,+0.02}_{\,-0.01}$ &  $-0.03\,^{\,+0.07}_{\,-0.07}\,^{\,+0.01}_{\,-0.01}$\\
$14.18 - 16.00$  & $148\pm13$ & $2.07 \pm 0.20 \pm 0.08$ & $0.08\,^{\,+0.28}_{\,-0.08}\,^{\,+0.02}_{\,-0.01}$ &  $-0.01\,^{\,+0.12}_{\,-0.06}\,^{\,+0.01}_{\,-0.01}$ \\
$16.00 - 18.00$  & $141\pm13$ & $1.77 \pm 0.18 \pm 0.09$ & $0.18\,^{\,+0.22}_{\,-0.14}\,^{\,+0.01}_{\,-0.04}$ &  $-0.09\,^{\,+0.07}_{\,-0.09}\,^{\,+0.02}_{\,-0.01}$ \\
$18.00 - 22.00$  & $114\pm13$ & $0.78 \pm 0.10 \pm 0.04$ & $0.14\,^{\,+0.31}_{\,-0.14}\,^{\,+0.01}_{\,-0.02}$ & $\phantom{-}0.02\,^{\,+0.11}_{\,-0.11}\,^{\,+0.01}_{\,-0.01}$\\
\hline
$1.00 - 6.00$ & $357\pm21$ & $2.41 \pm 0.17 \pm 0.14$ & $0.05\,^{\,+0.08}_{\,-0.05}\,^{\,+0.04}_{\,-0.02}$ & $\phantom{-}0.02\,^{\,+0.05}_{\,-0.03}\,^{\,+0.02}_{\,-0.01}$ \\
\end{tabular}
}
\end{table*}

\section{Systematic uncertainties} 

For the differential branching fraction measurement, the largest source of systematic uncertainty comes from an uncertainty of $\sim 4\%$ on the \decay{\Bp}{\Kp\jpsi} and \decay{\jpsi}{\mumu} branching fractions~\cite{PDG2012}. The systematic uncertainties are largely correlated between the \qsq bins. The uncertainties coming from the corrections used to calibrate the performance of the simulation to match that of the data are at the level of $1 - 2$\%. The uncertainties on these corrections are limited by the size of the \decay{\Dstarp}{\Dz (\to \Km\pip) \pip} and \decay{\jpsi}{\mumu} control samples that are used to estimate the particle identification and tracking performance in the data. The signal and background mass models are also explored as a source of possible systematic uncertainty. In the fit to the $\Kp\mumu$ invariant mass it is assumed that the signal line-shape is the same as that of the \decay{\Bp}{\Kp\jpsi} decay. In the simulation, small differences are seen in the \Bp mass resolution due to the different daughter kinematics between low and high \qsq. A 4\% variation of the mass resolution is considered as a source of uncertainty and the effect on the result found to be negligible.

For the extraction of $A_{\rm FB}$ and $F_{\rm H}$, the largest sources of uncertainty are associated with the event weights that are used to correct for the detector acceptance. The event weights are estimated from the simulation in $0.5\gev^{2}/c^{4}$ wide \qsq bins (driven by the size of the simulated event sample). At low \qsq, the acceptance variation can be large (at extreme values of $\cos\theta_{\ell}$) over the \qsq bin size. The order of the uncertainty associated with this binning is estimated by varying the event weights by half the difference between neighbouring \qsq bins and forms the dominant source of systematic uncertainty. The size of these effects on $A_{\rm FB}$ and $F_{\rm H}$ are at the level of $0.01 - 0.03$ and $0.01 - 0.05$ respectively, and are small compared to the statistical uncertainties. Variations of the background mass model are found to have a negligible impact on $A_{\rm FB}$ and $F_{\rm H}$. 

The background angular model is cross-checked by fitting a template to the angular distribution in the upper mass sideband and fixing this shape in the fit for $A_{\rm FB}$ and $F_{\rm H}$ in the signal mass window. This yields consistent results in every \qsq bin. Therefore, no systematic uncertainty is assigned to the background angular model.  Two further cross checks have been performed. Firstly, $A_{\rm FB}$ has been determined by counting the number of forward- and backward-going events, after subtracting the background. Secondly, $F_{\rm H}$ has been measured by fitting the $\left|\cos\theta_{\ell}\right|$ distribution, which is independent of $A_{\rm FB}$. Consistent results are found in both cases.

\section{Conclusions}

The measured values of $A_{\rm FB}$ and $F_{\rm H}$ are consistent with the SM expectations of no forward-backward asymmetry and $F_{\rm H} \sim 0$.  The differential branching fraction of the \decay{\Bp}{\Kp\mumu} decay is, however, consistently below the SM prediction at low \qsq. The results are in good agreement with, but statistically more precise than, previous measurements of $\deriv\BF/\deriv\qsq$ and $A_{\rm FB}$ from \babar~\cite{Aubert:2006vb, *:2012vw}, \belle~\cite{:2009zv} and CDF~\cite{Aaltonen:2011ja}. Integrating the differential branching fraction, over the full \qsq range, yields a total branching fraction of ${( 4.36 \pm 0.15 \pm 0.18 ) \times 10^{-7}}$, which is more precise than the current world average of $( 4.8 \pm 0.4 )\times 10^{-7}$~\cite{PDG2012}.

\section*{Acknowledgements}

\noindent We express our gratitude to our colleagues in the CERN accelerator
departments for the excellent performance of the LHC. We thank the
technical and administrative staff at CERN and at the LHCb institutes,
and acknowledge support from the National Agencies: CAPES, CNPq,
FAPERJ and FINEP (Brazil); CERN; NSFC (China); CNRS/IN2P3 (France);
BMBF, DFG, HGF and MPG (Germany); SFI (Ireland); INFN (Italy); FOM and
NWO (The Netherlands); SCSR (Poland); ANCS (Romania); MinES of Russia and
Rosatom (Russia); MICINN, XuntaGal and GENCAT (Spain); SNSF and SER
(Switzerland); NAS Ukraine (Ukraine); STFC (United Kingdom); NSF
(USA). We also acknowledge the support received from the ERC under FP7
and the Region Auvergne.

\addcontentsline{toc}{section}{References}
\bibliographystyle{LHCb}
\bibliography{main,rare,stat}

\end{document}